\documentclass[twocolumn,pre,superscriptaddress]{revtex4}
\usepackage{epsfig}

\usepackage{dcolumn}
\usepackage{amsmath}

\usepackage{graphicx}
\usepackage{rotating}
\usepackage{multirow}

\def\be{\begin{equation}}
\def\ee{\end{equation}}
\def\lsim{\raise0.3ex\hbox{$<$\kern-0.75em\raise-1.1ex\hbox{$\sim$}}}
\def\gsim{\raise0.3ex\hbox{$>$\kern-0.75em\raise-1.1ex\hbox{$\sim$}}}
\def\Tr{\mathop{\rm Tr}}

\def\PRL{{ Phys.\ Rev.\ Lett.\ }}

\def\EPJ{{ Eur.\ Phys.\ J }}

\begin{document}

\newlength{\figurewidth}
\ifdim\columnwidth<10.5cm
  \setlength{\figurewidth}{0.95\columnwidth}
\else
  \setlength{\figurewidth}{10cm}
\fi
\setlength{\parskip}{0pt}
\setlength{\tabcolsep}{6pt}
\setlength{\arraycolsep}{2pt}

\title{A Method to Find Community Structures Based on Information Centrality}
\medskip

\author{Santo Fortunato}

\vskip0.3cm

\affiliation{Fakult\"at f\"ur Physik, Universit\"at Bielefeld, 
D-33501 Bielefeld, Germany}

\author{Vito Latora}

\vskip0.3cm

\affiliation{Dipartimento di Fisica e Astronomia, Universit\`a di Catania
and INFN sezione di Catania, Via S. Sofia 64, 95123 Catania, Italy}

\author{Massimo Marchiori}

\vskip0.3cm

\affiliation{WSC and Lab. for Computer Science, Massachusetts Institute of
  Technology, USA}

\begin{abstract}
\noindent
Community structures are an important feature of many 
social, biological and technological networks. 
Here we study a variation on the method for detecting such 
communities proposed by Girvan and Newman and based 
on the idea of using centrality measures to 
define the community boundaries ( M. Girvan and  M. E. J. Newman, 
Community structure in social and biological networks 
Proc. Natl. Acad. Sci. USA  99, 7821-7826 (2002)). 
We develop an algorithm of hierarchical clustering that consists in 
finding and removing iteratively the edge with the highest 
information centrality. 
We test the algorithm on computer generated and real-world networks 
whose community structure is already known or has been studied 
by means of other methods. 
We show that our algorithm, although it runs to completion in a time $O(n^4)$, 
is very effective especially when the 
communities are very mixed and hardly detectable by the 
other methods. 
\end{abstract}

\maketitle

\vskip0.7cm

\section{Introduction}
Network analysis has revealed as a powerful approach to understand 
complex phenomena and organization in social, biological and 
technological systems 
\cite{wasserman,scott,mendesbook,strogatzreview,barabasireview}. 
In the framework of network analysis a given system is modeled as 
a graph in which the  nodes are the elements of the system, 
for instance the individuals in a social system, the neurons in a 
brain and the routers in the Internet, and the edges represent 
the interactions, social links, synapses and electric wirings 
respectively, between couples of elements.  
A lot of interest has been focused on the characterization of 
various structural and locational properties of the network  
\cite{wasserman,scott,mendesbook,strogatzreview,barabasireview}. 
Among the others, an important property common to many 
networks is the presence of {\it subgroups} or 
{\it community structures}. 
\\
For instance, in {\it social networks} some individuals 
can be part of a tightly connected 
group or of a closed social elite, others can be 
completely isolated, while some others 
may act as bridges between groups. 
The differences in the way that individuals are embedded in 
the structure of groups within the network can have important 
consequences on the behavior they are likely to practice. 
The division of the individuals of a social network 
into communities is a fundamental aspect of a social system. 
In fact, subgroups in social systems often have their own norms, 
orientations and subcultures, sometimes running counter to the 
official culture, and are the most important 
source of a person's identity \cite{scott}.
For this reason one of the main concerns, since the very 
beginning of social network analysis, has been the 
definition and the identification of subgroups of individuals 
within a network. And the first algorithms to find
community structures have been proposed in social network 
analysis. 
\\
Subgroups are also important to other networks. 
The presence of subgrouping in {\it biological} and {\it technological} 
networks may hinder important information on the functioning 
of the system, and can be relevant to understand the  
growth mechanisms of such networks.  
In fact, communities in the World-Wide-Web 
may represent pages on common topics, while 
community in cellular \cite{bio} and genetic 
networks \cite{huberman_genes} might represent 
functional modules \cite{ves_modularity}. 
For this reason, the techniques to find   
the substructures within a network provide a powerful tool 
for understanding the structure and the functioning of the 
network.
\\
In this paper we present a new method to discover 
community structures that uses the recently introduced 
{\it information centrality measure} 
\cite{lmcentrality,lminfrastructures},   
based on the concept of network global efficiency \cite{lm2,lm4}.  
The information centrality is here used to quantify 
the relevance of each of the edges in the network. 
The method consists in finding and removing the 
edges with the highest centrality score until 
the network breaks up into components. 

The paper is organized as follows. In Section \ref{definition} 
we review the definitions of cliques and cohesive 
subgroups and the standard methods for finding community 
structures in networks. In Section \ref{new} we propose the new 
method and describe its implementation.  
In Section \ref{test} we discuss the application of 
the algorithm to computer-generated networks 
for which there is already a knowledge and control on 
the existing subgroups. 
We show that the algorithm, although slower than the best 
methods on the market, can be extremely effective at 
discovering community structures, especially when the communities
are very mixed and hardly detectable. 
Finally in Section \ref{applications} we discuss a number of 
applications to real-world networks. 
In Section \ref{final} we present our conclusions.

\section{Definition of cohesive subgroups}
\label{definition} 
Social analysts were the first to formalize the idea 
of communities and to devise mathematical measures of the number 
and cohesion of communities. Here we review the most important 
definitions developed for social systems. For this reason 
the discussion of this section will be mainly in terms 
of social networks, although, as we will see in the 
following sections, the ideas of community structures 
applies as well to other networks.  
A {\it community}, or {\it cluster}, 
or {\it cohesive subgroup} is a subset of individuals  
among whom there are relatively strong, direct, intense ties. 
The starting point of all the definitions and measures is the concept 
of subgraph. A {\it subgraph} is any collection of nodes selected 
from the nodes of the whole graph, together with the 
edges connecting those nodes. A random sample of points in a graph 
representing a social system is for example a subgraph but it is not 
likely to correspond to any meaningful social group. 
The notion of a meaningful social group is based on the 
property of cohesion among the various members of the subgraph. 
However the cohesion of a subgraph can be quantified by using 
various different properties of the ties among subsets of nodes. 
The choice of a particular property instead of another depends on 
the researcher's decision that a particular mathematical criterion 
can be given a meaningful and useful sociological interpretation. 
The general aim is to define a meaningful social category by 
investigating the structural properties of the whole graph and 
finding the naturally existing communities into which the social 
network can be divided. 
\\
The literature on cohesive subgroups contains various ways 
to conceptualize the idea of subgroups in social networks. 
In particular, there are four main ideas that take into account four 
different structural properties \cite{wasserman}. 
The resulting four categories of cohesive subgroups  
are sorted in such a way that going from the first 
to the last one we weaken the properties that the subgroups 
have to fulfill. 
We briefly present these ideas for one-mode, non-directed, 
non-valued graphs.

{\bf 1) The mutuality of ties}. 
Cohesive subgroups based on the mutuality of ties require 
that all pairs of subgroup members choose each other. 
This idea is formalized in the definition of cliques. 
A {\it clique} is a maximal complete subgraph of three or more 
nodes, i.e. a subset of nodes all of which are adjacent 
to each other and there are no nodes that are also adjacent 
to all the members of the  clique. 

{\bf 2) The closeness or reachability} 
of the members of the subgroup.   
Since the definition of clique is rather strong and restrictive  
for real social networks, a number of extensions of the basic 
idea have been proposed. 
Cohesive subgroups based on reachability require that all 
the members are reachable from each other. 
The n-cliques extend the notion of cliques, weakening the requirement 
of adjacency among all the subgroup members. 
A {\it n-clique} is a maximal subgraph in which the largest 
geodesic distance between any two nodes is no greater than $n$. 
When $n=1$ we go back to the concept of clique. 2-cliques 
are subgraphs in which all nodes need not to be adjacent but 
are reachable through at most one intermediary. 
In 3-cliques all nodes are reachable  through at most two 
intermediaries, and so on. 
\\
A definition that will be important in the 
following of the paper is that of component. 
A {\it component} is the maximal connected subgraph, i.e. 
a subgraph in which there is a path between all pairs of 
nodes, while there is no path between a node in the 
subgraph and any node not in the subgraph.

{\bf 3) The frequency of ties} among members. 
This idea of cohesive subgroups is  based on restrictions 
on the minimum number of actors adjacent to each other in 
a subgroup. 
Whereas the concept of n-clique involves increasing 
the permissible path lengths, an alternative way to relax the strong 
assumption of cliques involves reducing the number of other 
nodes to which each node must be connected. 
A {\it k-plex} is a maximal subgraph containing $n$ nodes 
in which each node is adjacent to no fewer than $n-k$ nodes in 
the subgraph. Compared to n-clique analysis,  
k-plex analysis tends to find a relatively 
large number of smaller groups.

{\bf 4) The relative frequency of ties} among subgroup members 
compared to non-members. This idea of cohesive subgroups is 
different from the previous three because it is based on 
the comparison of ties within the subgroup to ties outside 
the subgroup \cite{alba}. 
In this way cohesive subgroups are seen as areas of relatively 
high density in the graph, parts that are locally denser 
than the field as a whole. 
The LS set is the simplest formal definition of a 
subgroup in this class.   
An {\it LS set} is a set of nodes $S$ such that any of its 
proper subsets (i.e. any possible subset of nodes that can be 
selected from the nodes in $S$) has more ties to its complement 
within $S$ than to the outside of $S$ \cite{seidman}. 
The fact that LS sets are related by containment implies that 
there is a hierarchy of LS sets in a graph. 
The definition of lambda sets extends that of LS sets,  
and is based on the concept of edge connectivity. 
The edge connectivity of a pair of nodes $i$ and $j$ 
is equal to the minimum number of edges that must be 
removed from the graph in order to leave no path 
between the two nodes. 
A set of nodes $S$ is a {\it lambda set} if any pair of nodes 
in $S$ has larger edge connectivity than any pair of nodes 
consisting of one node within $S$ and a node outside $S$ 
\cite{borgatti}. Lambda sets are based on the idea that a 
cohesive subgroup is relatively robust, namely it is hard 
to disconnect by the removal of edges. 
An alternative approach based on the same idea is to 
consider if there are edges in the graph which, if 
removed, would result in a disconnected structure. 
This approach is easy to implement into an algorithmic 
procedure and allows to develop {\it hierarchical clustering 
methods}. 
Such methods rank and remove the edges of the network in 
terms of their importance, where the edge importance can be 
defined in different ways as will be clear in a moment. 
By doing this repeatedly the network breaks iteratively 
into smaller and smaller components until it breaks 
into a collection of single non-connected nodes. 
The resulting hierarchical structure 
to clusters can be represented by  
{\it dendrograms}, or hierarchical trees, as the one 
reported in Fig. \ref{randdend}, showing the 
clusters produced at each step of the subdivision. 
\\
Recently, Girvan and Newman have considered two forms of edge 
betweenness to measure the edge importance: the shortest path 
betweenness and the random-walk betweenness \cite{newgirv1,newgirv2,note1}.
The edge shortest path betweenness extends to the edges    
the node betweenness proposed by Freeman \cite{freeman} 
as a centrality measure for the nodes, 
and is defined as the number of shortest 
paths between pairs of nodes that run through that edge 
\cite{newgirv1}.  
The random-walk betweenness does consider random walks 
connecting all couples of nodes instead of the shortest 
paths (random walks have also been used to quantify 
the similarities-dissimilarities between nearest-neighbouring 
nodes in other algorithms for finding communities 
\cite{zhou}). 
\\
The algorithms by Girvan and Newman at each step 
identify and remove the edges that are the most 
between couples of nodes, in the sense that they are responsible for 
connecting many pairs of nodes. 
The method for finding community structures that we present 
in this paper is a modification of the method by 
Girvan and Newman.  
In our method we propose to identify directly the edges that 
when removed mostly disrupt the network's ability 
in exchanging information among the nodes. 
In fact, instead of the edge betweenness, we adopt 
a measure of centrality, the information centrality 
$C^I$ \cite{lmcentrality,lminfrastructures}, 
based on the concept of efficient propagation of information 
over the network  \cite{lm2,lm4}. 
The information centrality has revealed as an interesting 
quantity to characterize the centrality of the nodes of 
a network, and gives different results from the 
betweenness centrality \cite{lmcentrality}. 
For this reason we think that it might be useful to 
develop an algorithm of hierarchical clustering based 
on the edges information centrality. 
\\
After having described the formal definitions of cohesive 
subgroups based on the relative frequency of ties, 
we need to give some methods for assessing the 
cohesiveness of the subgroups. This is especially important 
in hierarchical clustering methods where one obtains a hierarchy of 
community structures, from the original graph to the extreme case 
in which all the nodes are disconnected: in this case 
the number of communities depends on the level at which 
the graph is partitioned, and we therefore need a criterium 
to say at which point to stop. 
One of the first measures of how cohesive a subgroup is, 
was proposed in Ref. \cite{bock} and is just the ratio of the number 
of ties (or the average strength of ties for a valued graph) within 
a subgroup divided by the number of ties from the subgroup to nodes 
outside the subgroup. 
This measure was recently extended in Ref. \cite{newgirv2} 
by the measure of modularity that we will discuss in 
Section \ref{test} and which proves to be
successful to express the degree of cohesiveness of
the communities of many networks. This is why it was   
recently proposed in Ref. \cite{newmanfast} to adopt the modularity 
itself as the quantity to maximize so to identify the best 
community structure. 
The numerical implementation of this maximization 
allows to analyze very large networks because it 
can be performed in a time which is by far shorter 
than the time required by all the previous algorithms.

\section{Our method for finding communities}
\label{new}
The algorithm for finding structures we propose here 
makes use of a recently introduced centrality measure 
\cite{lmcentrality,lminfrastructures}, 
that is based on the concept of efficient propagation 
of information over the network \cite{lm2,lm4}.  
We assume that the network we want to analyze can be represented 
as a connected, non-directed, non-valued graph ${\bf G}$
of $N$ nodes and $K$ edges. However, 
the extension to non-symmetric and valued data does not present 
any special problem and will be considered in a forthcoming 
paper \cite{rosario}. The graph ${\bf G}$ is described 
by the adjacency matrix $\mathbf{a}$, a $N \times N$ matrix whose 
entry $a_{ij}$ is equal to 1 if $i$ and $j$ are adjacent and 0 
otherwise. Two nodes in the graphs are said adjacent if they 
are connected by an edge. 
The entries on the main diagonal are undefined, 
and for convenience they are set to be equal to 0. 
We now give some definition that will be useful in the following. 
A {\it walk} is an alternating sequence of 
nodes and edges, where each edge is linked to both 
the preceding and the succeeding node. A {\it path} linking 
two nodes $i$ and $j$ is a walk from $i$ to $j$ 
in which all points and edges are distinct: the length of the  
path is the number of edges traversed to get from $i$ to $j$.  
The shortest path, or {\it geodesic}, between $i$ and $j$ is any path 
from $i$ to $j$ containing the minimum number of edges.   
\\
In order to describe how efficiently the nodes of the network 
$\bf G$ exchange information we use the {\em network efficiency\/} 
$E$, a measure introduced in refs. \cite{lm2,lm4}. 
Such a variable is based on the assumption that 
the information/communication in a network  
travels along the shortest paths (geodesics), and that the efficiency 
$\epsilon_{ij}$ in the communication 
between two nodes $i$ and $j$ is equal to the   
inverse of the shortest path lenght $d_{ij}$.  
The {\it efficiency} of $\bf G$ is the average of $\epsilon_{ij}$:   
\begin{equation} 
\label{efficiency}
E[{\bf G}]=
\frac{ {{\sum_{{i \ne j\in {\bf G}}}} \epsilon_{ij}}  } {N(N-1)}
          = \frac{1}{N(N-1)}
{\sum_{{i \ne j\in {\bf G}}} \frac{1}{d_{ij}}}
\end{equation}
and measures the mean flow-rate of information over $\bf G$. 
The quantity $E[{\bf G}]$ varies in the range $[0,1]$, and 
is perfectly defined also in the case of non-connected graphs.   
In fact, when there is no path between $i$ and $j$, we assume 
$d_{ij}=+\infty$ and consistently $\epsilon_{ij}=0$. 
Such a property will be extremely important for our 
algorithm. 
\\
A measure of node centrality, 
the so called {\it information centrality}, 
based on the network efficiency, has been recently proposed 
\cite{lmcentrality}. The same measure can be used to quantify 
the importance of groups and classes 
\cite{lmcentrality,lminfrastructures}. 

Here we use such measure to quantify the importance of an 
edge of the graph {\bf G}.    
The information centrality $C^I_k$ of the edge $k$
is defined as the relative drop in the network 
efficiency caused by the removal of the edge 
from $\bf G$:  
\begin{equation} 
\label{IC}
C^I_k  =  \frac{\Delta E}{E} = 
                \frac{E[{\bf G}] - E[{\bf G}^{\prime}_k]}{E[{\bf G}]}
~~~~~~~~~k=1,...,K
\end{equation}
Here by ${\bf G}^{\prime}_k$ we indicate a  
graph with $N$ points and $K-1$ edges obtained 
by removing the edge $k$ from $\bf G$. 
Notice that this measure is perfectly defined also when 
${\bf G}^{\prime}_k $ is a non-connected graph. 
\\
The method for finding the hierarchy of cohesive subgroups 
in ${\bf G}$ consists in the iterative removal of the 
edges with the highest information centrality, until 
the system breaks up into components. 
We expect that the edges that lie between communities 
are those with the highest information centrality, 
while those inside communities have a low information 
centrality. 
The general form of the algorithm is the following: 
\begin{enumerate}
  \item Calculate the information centrality score for each of the 
        edges. 
  \item Remove the edge with the highest score.  
  \item Perform an analysis of the network's components.
  \item Go back to point 1 until all the edges are removed 
        and the system breaks up into N non-connnected nodes.  
\end{enumerate} 
As in the Girvan and Newman algorithms \cite{newgirv1,newgirv2},   
the recalculation of the information centrality scores 
every time after an edge as been removed appears to be an 
important aspect of the algorithm. 
We will discuss this point in Section \ref{applications}.  
The calculation of all the shortest paths, necessary to compute 
the efficiency of the network, can be performed with 
a breadth-first search algorithm in time $O(KN)$ \cite{newmanomn,brandes}. 
Then the calculation of the information centrality for all the 
edges takes a time $O(K^2N)$. This time is comparable to the time 
it takes to compute the random-walk betweenness for all the 
edges \cite{newgirv2}, but is longer than the time $O(KN)$ 
it takes to calculate the shortest paths betweenness 
for all the edges used in the method of Ref. \cite{newgirv1}. 
The algorithm repeats the calculation of all the 
information centralities for each edge removed, i.e. $K$ times. 
In conclusion, the entire community structure algorithm based 
on the information centrality can be completed in time $O(K^3N)$, 
or time $O(N^4)$ for a sparse graph. Although, as we will show  
in Section \ref{test}, the algorithm can be in some cases 
better in finding community structures than the algorithm based 
on shorthest path betweenness, for its poor performance 
it can be used only for graphs with up to a thousand of nodes. 
For extremely large networks the best algorithm to be used 
is the one proposed in Ref. \cite{newmanfast} and based on the 
maximization of the modularity that runs in time $O(KN)$ or 
$O(N^2)$ on a sparse graph, or the one proposed in  
Ref. \cite{huberman} based on the notion of voltage drops across 
the network and running in time $O(K+N)$.

\section{Testing the method on computer generated networks}
\label{test}
\begin{figure}[htb]
\begin{center}
\resizebox{\figurewidth}{!}{\includegraphics[angle=90]{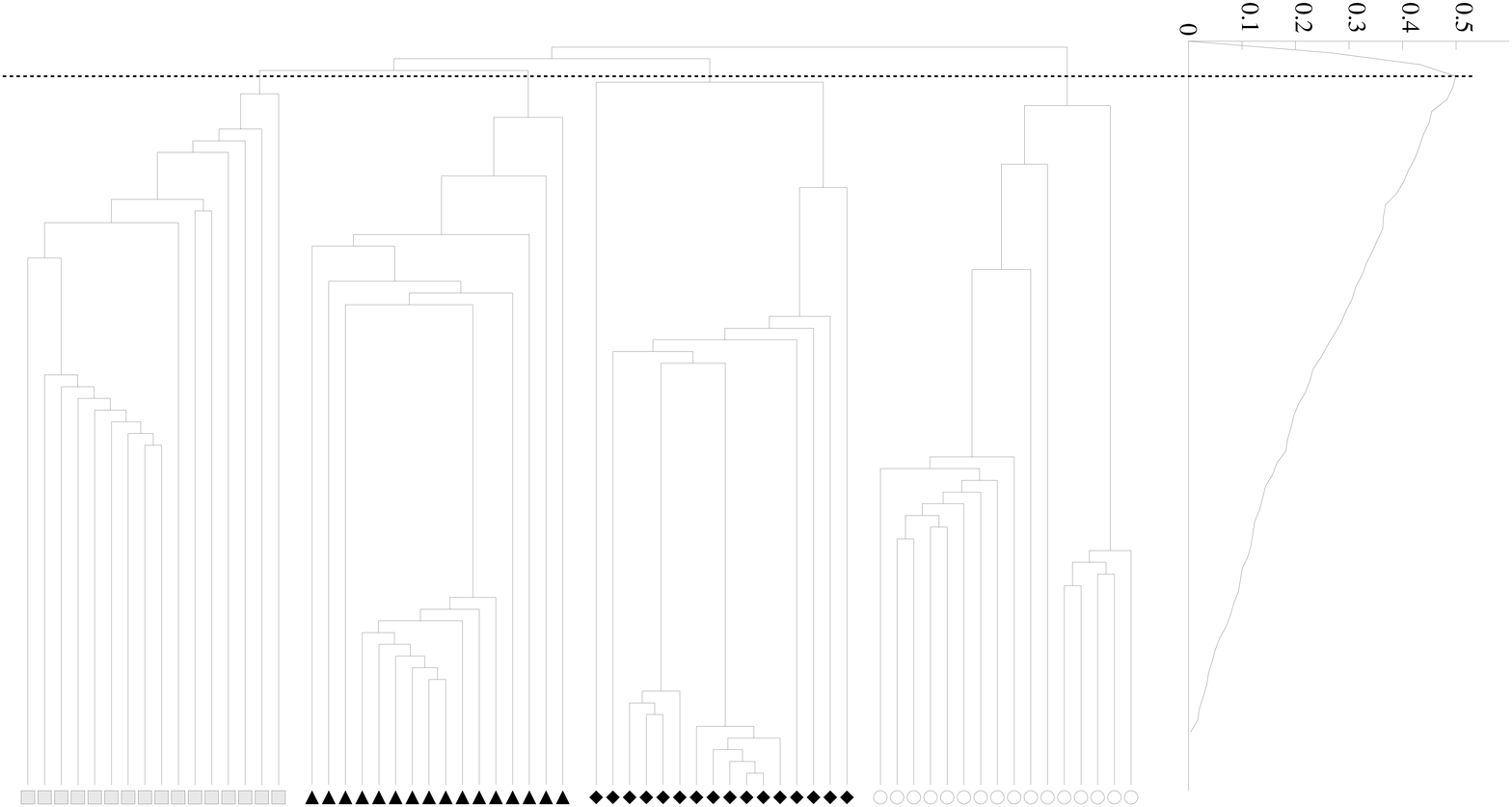}}
\end{center}
\caption{Dendrogram of the communities 
found by applying our algorithm to a computer generated random 
graph with 64 vertices and 256 edges. 
The random graph has been obtained by dividing the nodes 
into 4 groups of 16 nodes each (respectively empty 
circles, full circles, triangles and squares) 
and considering 
$z_{in}=6$, $z_{out}=2$ (see text). 
In the top panel the value of Q corresponding to the various divisions 
of the dendrogram is reported.}  
\label{randdend}
\end{figure}

We first applied our algorithm to computer generated networks, i.e.
random graphs constructed in such a way that 
they have a well defined
community structure. All graphs have the same number of nodes,
128, and the same number of edges, 1024. The nodes are divided 
into four classes, which are the groups 1-32, 33-64, 65-96 and 97-128.
We fixed to 16 the average number of edges per node, and 
we label the edges according to whether they connect members of the same 
group or not. The mixing between the classes is introduced by tuning 
the average number of edges connecting nodes belonging to 
different classes.
From a generic vertex of the graph
we have on average $z_{in}$ edges which join it to other vertices of its group
and $z_{out}$ edges connecting it to vertices of the other groups.
The two numbers are not independent, as we must of course have 
$z_{in}+z_{out}=16$. We remark that this is the same set of graphs that 
Newman \cite{newmanfast} and previously 
Girvan and Newman \cite{newgirv1,newgirv2} 
have used to test their algorithms. In this way we are able
to compare directly the role of edge betweenness and edge 
information centrality in determining
the community structure.
As a practical example we show in Fig. \ref{randdend} 
the dendrogram corresponding to the analysis with our method of a graph 
of this type, where for illustration purposes we take a smaller network
with 64 nodes and 8 edges per node. Here, $z_{in}=6$ and $z_{out}=8-z_{in}=2$, 
i.e. the network is strongly clustered. The algorithm produces a 
hierarchy of subdivisions of the network: from a single component to 
N isolated nodes, going from top to bottom in the dendrogram (left to right in
the figure).  
To know which of the divisions is the best one for a given
network i.e. where we have to cut the hierarchical tree, we need to 
use a measure of the cohesiveness of the communities.  
The first measure of how cohesive a subgroup is, 
was proposed in Ref. \cite{bock}. 
If there are $N$ nodes in the graph $\bf G$ and $N_S$ nodes in the 
subgroup ${\bf S}$, the cohesiveness of subgroup ${\bf S}$ can be 
defined as the ratio of the number of ties within subgroup 
${\bf S}$ divided by the number of ties from ${\bf S}$  to outsiders : 
\begin{equation} 
\frac{ \sum_{i \in S} \sum_{j \in S}  a_{ij} }
     { \sum_{i \in S} \sum_{j \notin S} a_{ij} }
\end{equation}
This measure was recently extended by Girvan and Newman in 
Ref. \cite{newgirv2} into the measure of modularity,  
that allows to consider more than a group at the same time and tell 
us how good a subdivision of ${\bf G}$ in $n$ subgroups is.   

\begin{figure}[htb]
\begin{center}
\resizebox{\figurewidth}{!}{\includegraphics[angle=0]{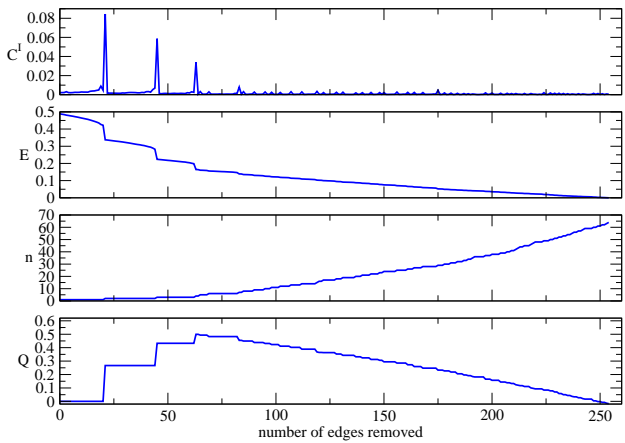}}
\end{center}
\caption{Information centrality $C^I$ of the edge removed, 
global efficiency $E$, number of components $n$ and modularity $Q$ 
for the resulting graph as a function of the number of 
edges removed.  
}  
\label{randeff}
\end{figure}

The modularity $Q$ is defined in the following way. 
Let us suppose that we want to test the goodness of 
a subdivision of the network in $n$ well defined communities. 
We expect that a good split is obtained if most of the edges 
fall inside the communities, with comparatively few edges 
joining the communities to each other. For this purpose one introduces
a $n\times n$ symmetric 
matrix $\mathbf{e}$ whose element $e_{ij}$ is the fraction of all edges 
in the network that link vertices in community~$i$ to vertices in 
community~$j$ \cite{note2}.   
The trace of this matrix $\Tr \mathbf{e} =\sum_i e_{ii}$ gives 
the fraction of edges 
in the network that connect vertices in the same community.
To try just to maximize the value of the trace does not help
because by considering the whole network as a single community
we would get the maximal value $1$ without doing any subdivision at all.
Therefore we further define the row (or column) sums $a_i=\sum_j e_{ij}$, which
represent the fraction of edges that connect to vertices in community~$i$.
If the network is such that the probability to have an edge 
between two sites is the same regardless of
their eventual belonging to the same community 
(random network), we would have $e_{ij}=a_i a_j$.  The 
modularity is defined as
\begin{equation}
Q = \sum_i \bigl( e_{ii} - a_i^2 \bigr) = \Tr {\mathbf{e}} - ||
{\mathbf{e}^2} ||
\label{defsq}
\end{equation}
where $ || {\mathbf{e}^2} ||$ indicates the sum of the elements of the matrix
$\mathbf{e}^2$. 
This quantity then measures the 
degree of correlation between the probability of having an edge joining
two sites and the fact that the sites belong to the same community.
It now makes sense to look for high values of $Q$. In fact, if we take
the whole network as a single community, we get $Q=0$ and we can easily get
higher values by choosing subdivisions in more than just a single class.
Values approaching $Q=1$, which is the maximum, indicate strong community
structure; on the other hand, for a random network $Q=0$.
The expression (\ref{defsq}) is not normalized, so that $Q$
will not reach a value
of~1, even on a perfectly mixed network.
For networks with an appreciable subdivision in classes, $Q$  
usually falls in the range from about $0.2$ to~$0.7$.  

In Fig. \ref{randdend} we plot the $Q$ corresponding to the 
classes we determined after each split. 
The x-coordinate represents the number of steps of the algorithm which end
with a split of the network (or of one of its components, if the network
is not connected). We remark that, since $Q$ is always calculated 
by using the full network, $Q$ can only vary if, after the remotion
of one edge, the number of components of the network changes, otherwise
it keeps the value corresponding to the last subdivision.
To take for the $x$-coordinate the number of removed edges
would result in a plot with many intervals where $Q$ stays constant and, 
even if that would not affect our description, we do not consider it
appropriate for a presentation.

The plot presents a single peak, which exactly corresponds to the
splitting of the network into the four groups. This means 
that the algorithm succeeds in identifying the four classes. 
The height of the peak is 0.499, which indicates that the network 
is indeed highly clustered.
\\
In Fig. \ref{randeff} we show the details of the calculation. 
We plot the information centrality $C^I$ of the edge removed, 
the global efficiency $E$, the number of components $n$ of the 
resulting graph and the value of $Q$ as a function of the number 
of removed edges, 
i.e. as a function of the iterations of the algorithm. 
Each time we remove an edge with a high 
information centrality score, i.e. each time there is a sharp drop in 
the network efficiency, we also observe a sharp increase in the 
modularity. The height of the three main peaks  
in $C^I$ is roughly proportional to the corresponding variations
of $Q$. The correlation between $C^I$ and $Q$ is non-trivial, 
but we can give the following simple argument to explain it.
Suppose that after the removal of an edge we get a split of the component
$A$ in two classes, say $A_1$ and $A_2$. We indicate with 
$I_{A_1}$, $I_{A_2}$, $I_{A}$ the number of edges joining pairs of vertices 
within $A_1$, $A_2$ and $A$, respectively. Furthermore, let us denote with
$m_{A_1}$, $m_{A_2}$, $m_A$ the sum of the vertex degrees of all the vertices
of $A_1$, $A_2$ and $A$. According to Eq. \ref{defsq},
the modularity $Q_b$ before the split is

\begin{equation}
Q_b \,\sim\, \frac{I_A}{K}-(\frac{m_{A}}{2K})^2, 
\label{ex1}
\end{equation}

where $K$ is the total number of edges of the network.
Notice that $I_A/K$ is exactly $e_{AA}$ of Eq. \ref{defsq}
and $m_A/2K$ roughly $a_A$ (with $i=A$).
On the other hand, after the split, we get the modularity

\begin{equation}
Q_a \,\sim\, \frac{I_{A_1}+I_{A_2}}{K}-(\frac{m_{A_1}}{2K})^2-(\frac{m_{A_2}}{2K})^2. 
\label{ex2}
\end{equation}

As just a few edges keep $A_1$ and $A_2$ together in $A$,
$m_A$ is approximately given by $m_{A_1}+m_{A_2}$.
So, we come to the following expression for the modularity variation 
${\Delta}Q$ after the split:

\begin{equation}
{\Delta}Q=Q_a-Q_b \,\sim\, \frac{I_{A_1}+I_{A_2}-I_A}{K}-\frac{m_{A_1}\,m_{A_2}}{2K^2}. 
\label{ex3}
\end{equation}

\begin{figure}[htb]
  \epsfig{file=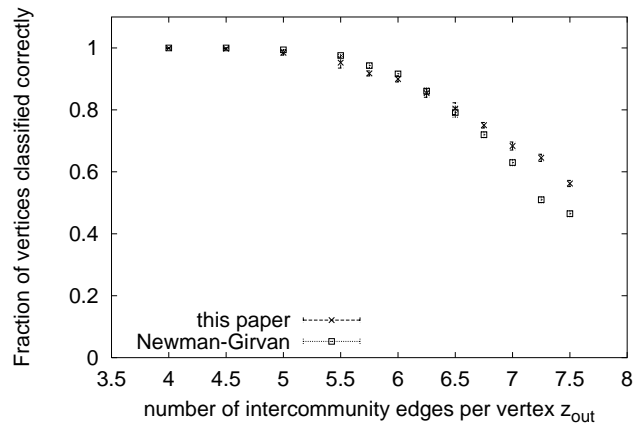,width=8.5cm}
  \caption{\label{randcompare}{Average fraction of correctly identified vertices
as a function of $z_{out}$. Each point represents an average
over 100 to 500 graphs. The comparison with the analogous results of 
Girvan and Newman shows that our algorithm performs better when the communities are 
very mixed and hardly detectable.}}
\end{figure}

The first term on the r.h.s. of Eq. \ref{ex3} is small, because 
$I_A\,{\sim}\,I_{A_1}+I_{A_2}$, so the dominant term is the second one, which 
is proportional to the product $m_{A_1}\,m_{A_2}$. On sparse graphs like those
we are dealing with here, $m_{A_1}\,m_{A_2}$ is roughly proportional to 
the number of vertex pairs with a vertex in $A_1$ and the other in $A_2$.
This number of pairs equals the number of paths going from $A_1$ to $A_2$,
which after the split are of infinite length and
give a vanishing contribution to the global efficiency
of the network. The variation of the information
centrality is then due to those paths, so it is proportional to ${\Delta}Q$, as
we find numerically.
\\
Our aim is of course to test how the algorithm works 
for many different networks, 
and this is accomplished by considering many different realizations of the 
same graph and checking how many vertices are correctly 
classified in each case. We analyzed our artificial networks 
for various values of $z_{out}$, ranging from $4$ to $7.5$, with a step of
$0.25$. We did not do a quantitative 
analysis of the interval $0<z_{out}<4$ because there
the algorithm always finds the right classes
(more than $99\%$ of successful attempts).
For each value of $z_{out}$ we produced from 100 to 500 samples, and calculated 
the average fraction of nodes which end up in their natural group.
We plot such averages in Fig. \ref{randcompare} 
as a function of $z_{out}$.
In the same plot we report the results obtained by using the algorithm 
of Girvan and Newman on the same network. We see that in the sector 
$[4,6]$ the two algorithms perform equally well; the algorithm of Girvan and
Newman seems to lead in some cases to slightly better results but they 
are compatible with ours within errors except eventually for $z_{out}=5.75$.
This is also the region of values of $z_{out}$ which corresponds to networks with
a clear community structure. In the sector $[6,7.5]$, where the communities
are very mixed and hardly detectable, both algorithms start 
inevitably to fail, but our algorithm clearly performs better. 
In $[7,7.5]$ our results are even better 
than the ones obtained through the modularity-based algorithm 
recently proposed by Newman \cite{newmanfast}. 
These results may justify the extra price in terms of CPU time that we have 
to pay if we choose to adopt the algorithm based on the information 
centrality. 
As far as the modularity is concerned, we passed from peak values of 
about 0.65 for the lowest $z_{out}$ we have taken (2) to about 0.25 for the most mixed
cases ($z_{out}=7.5$).

\begin{figure}[htb]
  \epsfig{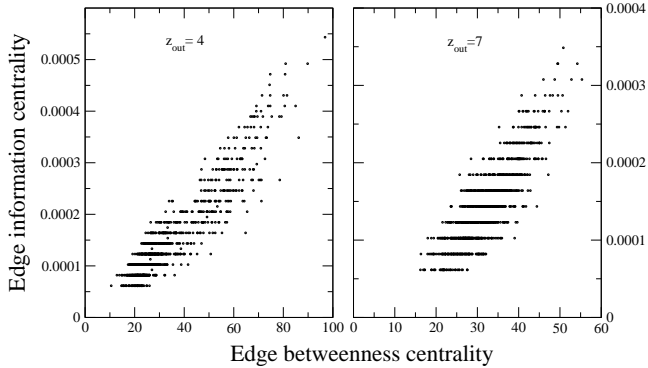}
  \caption{\label{infvsbet}{Correlation between edge information centrality 
and betweenness centrality. Each point of the scatter plot refers to 
an edge of an artificially generated network with 128 nodes and 1024 edges. 
We consider the two values $z_{out}=4$ and  $z_{out}=7$, respectively 
representing a case in which the communities are clearly 
separated and a case in which the 
communities are mixed and hardly detectable. 
}}
\end{figure}

As a further evidence of the similarities and 
differences between edge information  
and betweenness centrality we report in Fig. \ref{infvsbet} a scatter plot 
of the two measures for each of the 1024 edges of the initial 
network, i.e. before we start the first iteration of the edge removal 
process. The figure shows that, as expected, the two measures are 
correlated, although there are some important differences. 
In particular we notice that the edges with the higher 
information are not always those with the higher betweenness.  
This is more evident when the communities are mixed and hardly detectable. 
For instance in the case 
$z_{out}=7$ the edge with the largest information, i.e. the one that 
will be removed by our algorithm is not the one with the largest betweenness.  

\section{Applications to real networks}
\label{applications}
After the first experiments on artificial networks, we can say that the
algorithm seems promising. However, if our method is any good, it must 
work as well for real networks, which actually represent the 
systems we are mostly interested in. 
We present here the
analysis of four networks, although we analyzed
more. The first three of them, i.e. the 
Zachary's karate club, the network of the 
American college football teams and the food web 
of the Chesapeake Bay, have also been studied 
by other authors, with other hierarchical clustering 
methods. 
In this way we can better understand what the differences 
between the various approaches are. 
The last network studied represents the interactions amongst a group of 
20 monkeys. 

\subsection{Zachary's karate club}

\begin{figure}[htb]
  \begin{center}
    \epsfig{file=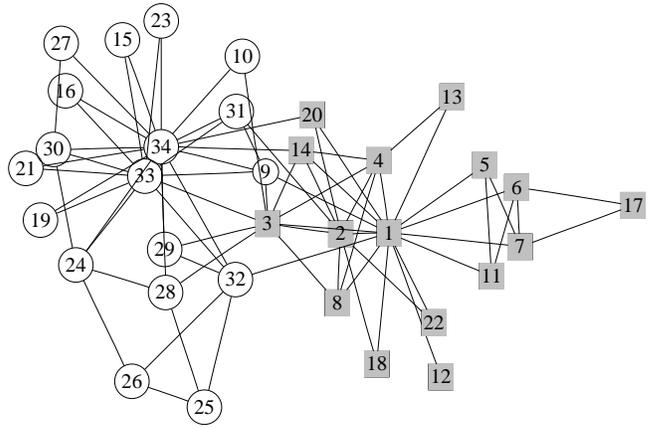,width=8.5cm}
    \caption{\label{zac}{The karate club network of Zachary (figure taken from
        Girvan and Newman \cite{newgirv2}).}}
  \end{center}
\end{figure}

\begin{figure}[htb]
  \begin{center}
    \epsfig{file=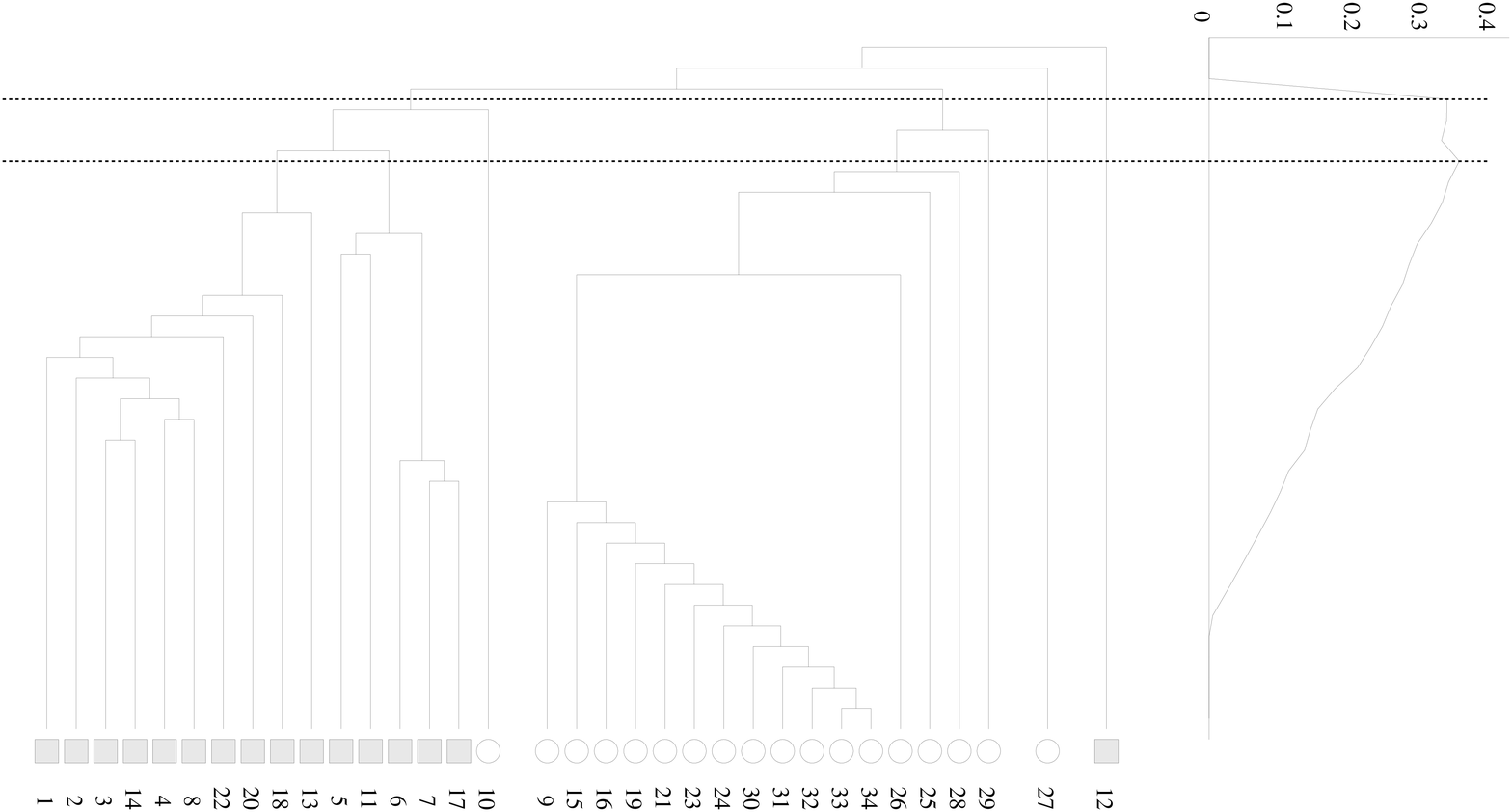,angle=90,width=9cm}
    \caption{\label{zacdend}{Dendrogram of the communities of the karate club. Initially one has
the split of two loosely bound nodes, $12$ and $27$, from the rest of the
network. After that the two
communities, with the exception of node $10$ (and of the two
 above-mentioned nodes), are correctly identified. The separation of the two
 communities corresponds to a peak in the modularity $Q$.}}
  \end{center}
\end{figure}

\begin{figure}[htb]
  \begin{center}
    \epsfig{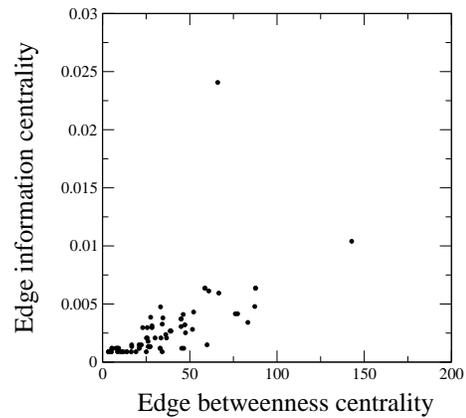}
    \caption{\label{karateinfvsbet}{Correlation between edge information 
centrality and betweenness centrality for the karate club network. 
Each point of the scatter plot refers to an edge of the network.}}
  \end{center}
\end{figure}

The first example we considered is the famous karate club network 
analyzed by Zachary \cite{zachary}. It consists of 34 persons 
(78 edges) whose mutual friendship relationships 
have been carefully investigated over a period of two years. 
Due to contrasts between a teacher and the administrator of the club, 
the club split into two smaller ones. 
The questions we want to answer are the following: 
Is it possible, by studying the network community 
structures before the network splitting, to predict the behavior of the 
network and in particular to identify the two classes ? 
Moreover, according to the network structure will a possible conflict 
most likely involve two factions or multiple groups ? 
The network is presented in Fig. \ref{zac}, where the squares and the circles 
label the members of the two groups.  
The results of our analysis are illustrated in the dendrogram of 
Fig. \ref{zacdend}.  
\\
The first edge which gets removed is the one linking node 12 
to the rest of the network. This edge corresponds to the edge between 
node 12 and node 1, an edge having the largest information centrality 
(0.024) and a medium value of betweenness (66) as shown in 
the scatter plot reported in Fig. \ref{karateinfvsbet}. Notice also that the 
edge with the highest betweenness (142.79) is the edge connecting 
node 1 with node 32. 
The removal of the first edge then leads to the
isolation of node 12. This is a feature that we encountered other times 
in our analyses. The early separation of a single node or of a small group  
is due to the fact that a system often looses more efficiency   
because of such splits than through the removal of 
intercommunities edges.

\begin{figure}[htb]
  \begin{center}
    \epsfig{file=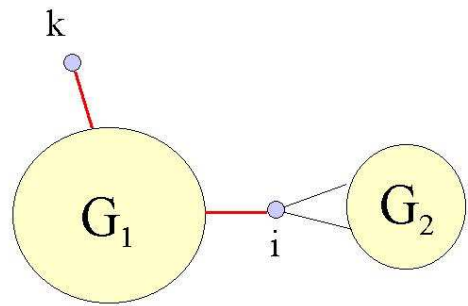,angle=0,width=9cm}
    \caption{\label{esempio}{A graph ${\bf G}$ composed by a node $k$ and 
two cohesive subgroups, ${\bf G_1}$ and ${\bf G_2}$, 
connected by node $i$. }}
  \end{center}
\end{figure}

To see why this is so, let us consider the simple example of 
Fig. \ref{esempio}, describing a network ${\bf G}$ with $N$ nodes 
composed by two cohesive subgroups, namely ${\bf G_1}$ with $N_1$ nodes, 
and ${\bf G_2}$ with $N_2$ nodes ($N_1 \sim N_2 \ll 1$), 
and by the two nodes $k$, which is joined to the network via a single edge 
(like node 12 in the karate club) and $i$, bridging  
${\bf G_1}$ to ${\bf G_2}$. 
In such a case the separation of the node $k$ 
leads to a decrease of efficiency 
proportional to the number of remaining nodes, i.e. 
$\Delta{E}_{k-split}\,\propto\,O(N)$.  
In fact, because of the single edge, the shortest paths between pairs 
of nodes different from $k$ are not affected by the removal of the edge, 
so the only contributions come from the
paths from $k$ to the rest of the network, which are 
$N-1$. 
On the other hand, the removal of the edge linking $i$ to 
${\bf G_1}$ influences the lengths of $N_1 \times N_2$ shortest paths, 
so that $\Delta{E}_{int-comm}\propto\,O(N^2)$. 
In such a case, the edge standing between the two communities 
${\bf G_1}$ and ${\bf G_2}$ will be the first one to be removed. 
But this is not always the case, since a simple modification of 
the network considered in the figure would lead to a different result. In 
fact, if we now suppose that node $i$ is connected to ${\bf G_1}$ 
through two edges (as for the connection between node $i$ and 
${\bf G_2}$) instead of a single one, then the algorithm 
will see the graph composed by  ${\bf G_1}$,  ${\bf G_2}$ and 
$i$ as a more cohesive structure than before and 
the first edge to be removed will be the one connecting 
$k$ to ${\bf G_1}$. 
\\
Going back to the dendrogram of Fig. \ref{zacdend}, we see that 
after node $12$ is removed from the network of the karate club, 
also the loosely bound node $27$ (just two edges) isolates from the rest. 
The third split finally separates the two big groups. At this stage 
we have four components, 
two isolated nodes ($12$ and $27$) and two larger groups which are homogeneous
except node $10$ which is misclassified (curiously enough,
this node is also misclassified by the fast algorithm   
of Newman \cite{newmanfast}).
The separation of the four above mentioned clusters corresponds to 
a peak in the plot of $Q$. However there is a second higher peak which
is obtained for a split of the network into seven communities. This 
double peak structure is present as well in the $Q$-plot of the 
Girvan-Newman analysis \cite{newgirv1,newgirv2}. 

\begin{figure}[htb]
\begin{center}
\resizebox{\figurewidth}{!}{\includegraphics[angle=0]{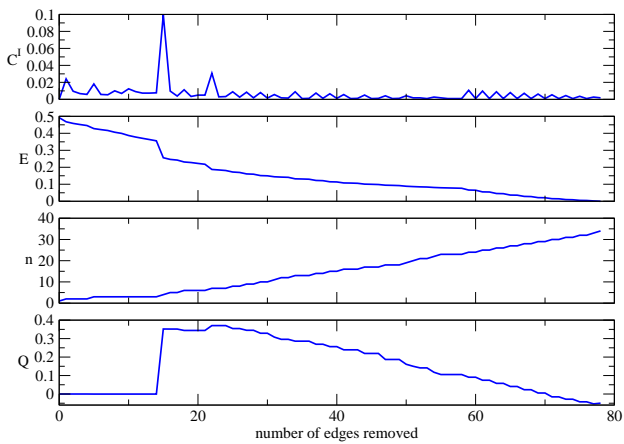}}
\end{center}
\caption{Information centrality $C^I$ of the edge removed, 
global efficiency $E$, number of components $n$ and value of $Q$ 
for the resulting graph as a function of the number of 
edges removed for the karate club network.}  
\label{karateeff}
\end{figure}

\begin{figure}[htb]
  \begin{center}
    \epsfig{file=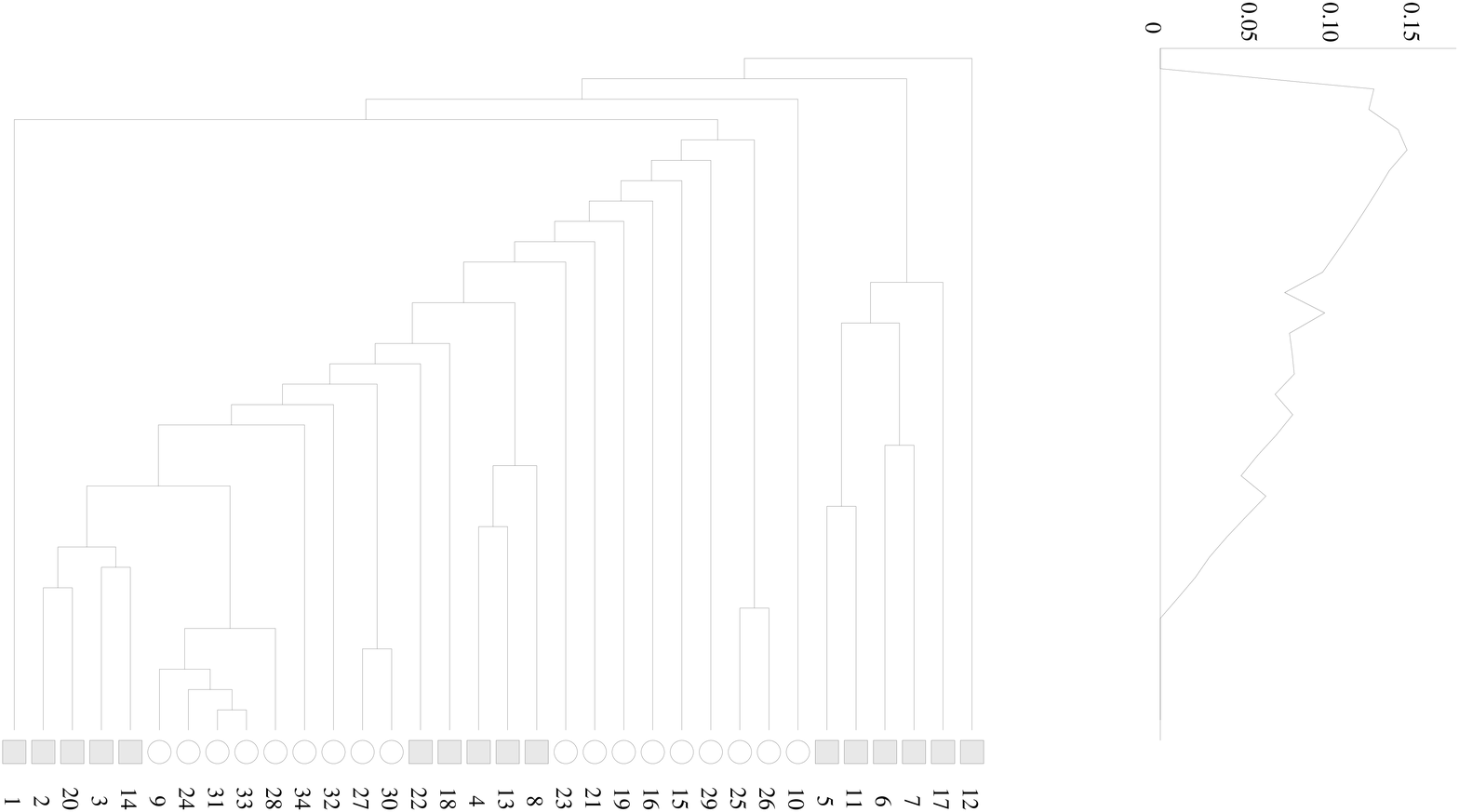,angle=90,width=9cm}
    \caption{\label{karate_norec}{Dendrogram of the communities of the karate
        club obtained by our method if we calculate the information
        centrality according to the 
        initial structure of the network. 
This version of the algorithm fails to
        detect the communities.}}
  \end{center}
\end{figure}

As for the computer generated networks, we report in  
Fig. \ref{karateeff} the information centrality $C^I$ of the edge removed, 
the global efficiency $E$, the number of components $n$ of the 
resulting graph and the value of $Q$ as a function of the number 
of edges removed from the network of the karate club. 
The figure is analogous to Fig. \ref{randeff}. 
We observe again a correlation between the peaks of
$C^I$ and the jumps of $Q$ (here we have two). 
Moreover, like in the previous case, the
absolute maximum of $Q$ corresponds to the lower of the two
peaks of $C^I$.
\\
We remind that the variation of the efficiency 
corresponding to the remotion of one edge is calculated by taking into account
the structure of the network at the current stage, i.e. without 
considering the edges which were eliminated in the previous steps. 
For the algorithm of Girvan and Newman this condition of recalculation
turns out to be crucial, because
removing the edges according to the (decreasing) values of the betweenness 
as calculated from the original configuration of the network
leads to very poor results. We wanted to check whether this 
is also true for our method. Indeed, Fig. \ref{karate_norec} clearly
shows that this is the case: the dendrogram does not reveal the 
real splitting of the network into the two classes, which instead 
look quite mixed up, and the modularity, whose values are quite low all over,
presents a rather flat profile.
%
%

\subsection{Network of the American college football teams} 

\begin{figure*}[htb]
  \begin{center}
\resizebox{\textwidth}{!}{\includegraphics{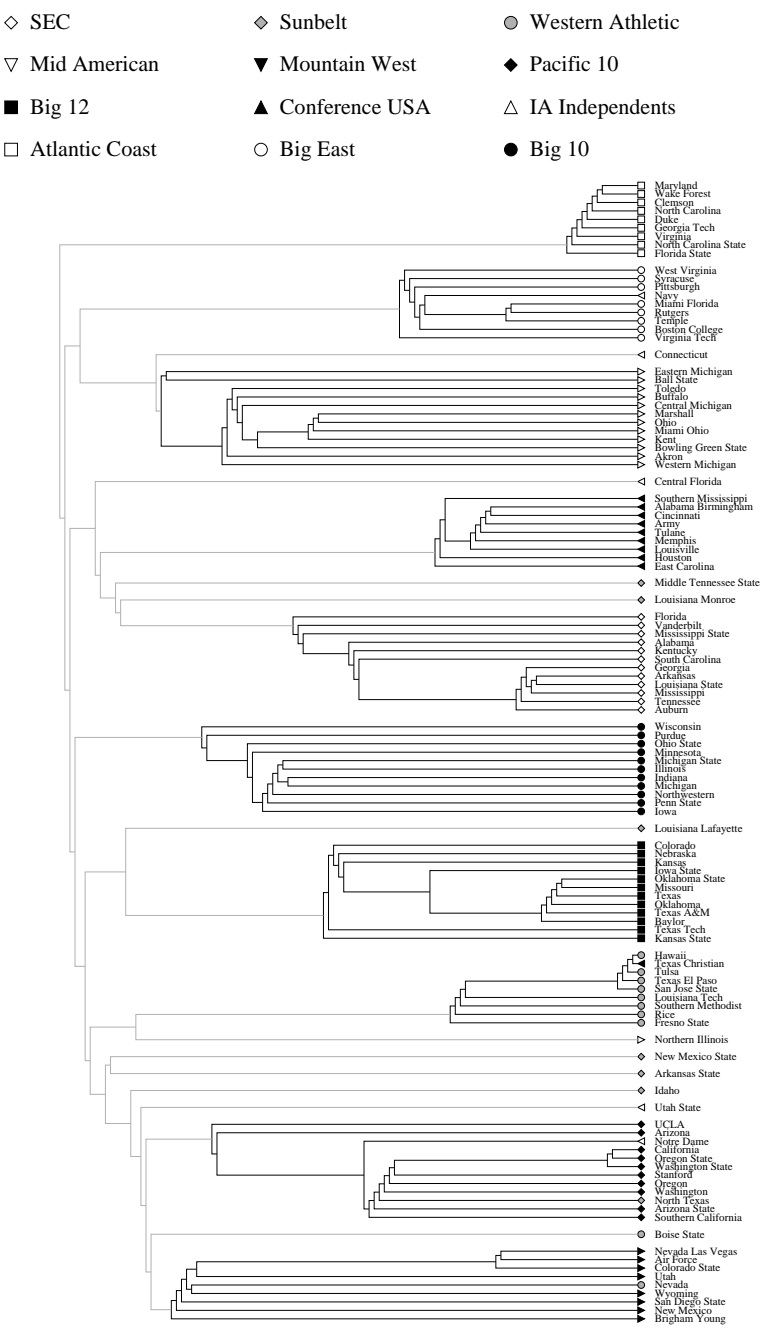}}
    \caption{\label{collegedend}{Dendrogram of the communities found in the
        college football network. At some
        stage we get to the highlighted structure, which shows a split into 
        ten main groups and few isolated nodes. The ten groups coincide, up to
        eventually a team, with
        ten of the conferences listed to the right of the figure.}}
  \end{center}
\end{figure*}

The second network we have investigated is the college football network,
representing the schedule of games between American college football
teams in a season. The teams are divided into well known "conferences",
which are the communities, with a higher number of games between
members of the same conference than between teams of different conferences.
There are altogether eleven conferences plus few other teams
which do not belong to any conference. 
Fig. \ref{collegedend} shows the dendrogram we have derived with our method.
The pattern of the modularity looks similar to the one we have shown for the
karate club, and it again presents two peaks, the higher of which reaches
the value $Q=0.485$. 
The corresponding subdivision of the network is the one we highlighted in the 
figure. We identify ten groups which coincide with 
ten conferences (either exactly or up to a team). 
The teams labeled as Sunbelt
are not recognized as belonging to the same group. This group is
misclassified as well
in the analysis of Girvan and Newman. There is however a reason
for that, namely the fact that 
the Sunbelt teams played basically the same number of games against
Western Athletic teams as they did among themselves. 
The independent teams (labeled as
IA Independent) show indeed no relationship to each other nor to 
a particular conference and they appear as truly independent nodes.

\subsection{Food Webs}

\begin{figure}[htb]
  \begin{center}
    \epsfig{file=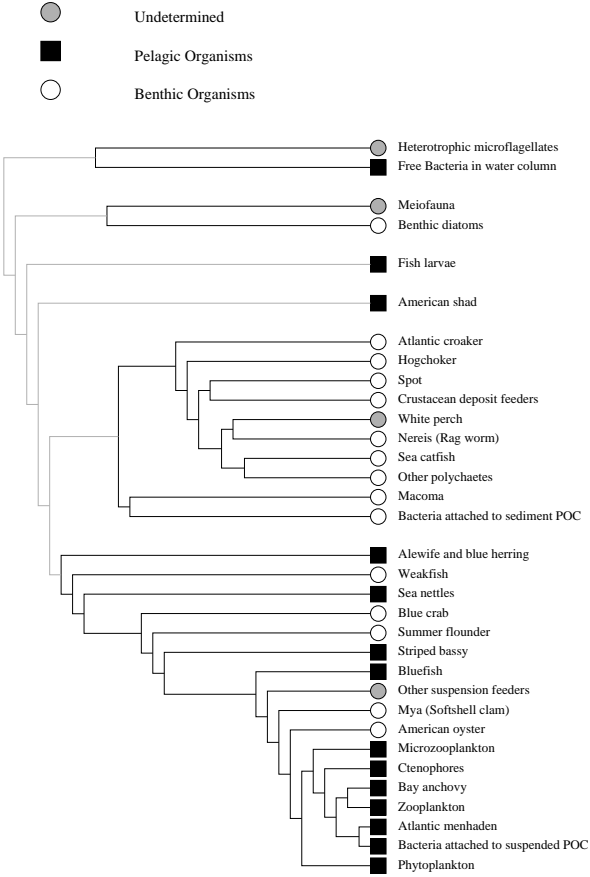,angle=90,width=8cm}
    \caption{\label{chesbay}{Dendrogram of the communities of the 
        Chesapeake Bay food web. The modularity peaks for the highlighted
        partition of the network. The two largest clusters are quite 
        homogeneous, reflecting approximately the division between 
      pelagic and benthic organisms.}}
  \end{center}
\end{figure}

We have also applied our algorithm to several food webs.
Here we mainly discuss the analysis of the food web of marine organisms
living in the Chesapeake Bay, which is situated on the Atlantic
coast of the United States. This special ecosystem
was originally studied by
Baird and Ulanowicz \cite{baird}, who carefully investigated the trophic 
relationships (i.e. the predatory interactions) between the 33 most important taxa,
which are the vertices of the network;
a taxon is a species or a group of species.
Baird and Ulanowicz studied the exchanges of carbon 
among the taxa and in this way they compiled the matrix
of their thophic relationships, specifying the percentage of
carbon assimilated in each interaction. 
So, strictly speaking, the network is directed ($A$ feeds on $B$ but the
opposite is not true), and valued (due to the different
percentages of carbon exchanged in the interactions); nevertheless, we took
it non-directed and non-valued, following
Girvan and Newman \cite{newgirv1}. 
Fig. \ref{chesbay} shows our analysis, which is quite similar to the
analysis of Ref. \cite{newgirv1}; the optimal split (peak of the modularity)
is obtained for a separation in two large classes and four small ones.
One of the big groups, with a few exceptions, contains  
pelagic organisms (which live near the surface or at middle depths),
the other one mainly benthic organisms (which live near the bottom).
Our classification of the taxa thus favours the habitat
versus the trophic levels, in contrast to other methods
used to study food webs.
We must be careful, however. On the one hand the mixed pattern 
of Fig. \ref{chesbay} suggests that one should probably take into 
account other criteria as well. On the other hand our analysis 
of a similar food web, relative to the seagrass ecosystem of
St. Marks National Wildlife Refuge \cite{baird2}, shows different results.
This network is larger than the previous one (48 vertices versus 33) and
has several species in common with the ecosystem of the 
Chesapeake Bay. The presence of terrestrial species and birds
enlarges the variety of possible habitats and the spectrum of the trophic
levels; the latter allowed to identify in \cite{baird2} five clusters of taxa.
Nevertheless, our study did not reveal 
any particular subdivision of the species. Repeating the analysis
with the algorithm of Girvan and Newman led essentially to the same results.
We had similar problems by analyzing other food webs;
the reason may be the fact that these networks often contain 
many edges, and our algorithm is probably not suitable
for the analysis of dense graphs.

\subsection{Primate Network}

\begin{figure}[htb]
  \begin{center}
    \epsfig{file=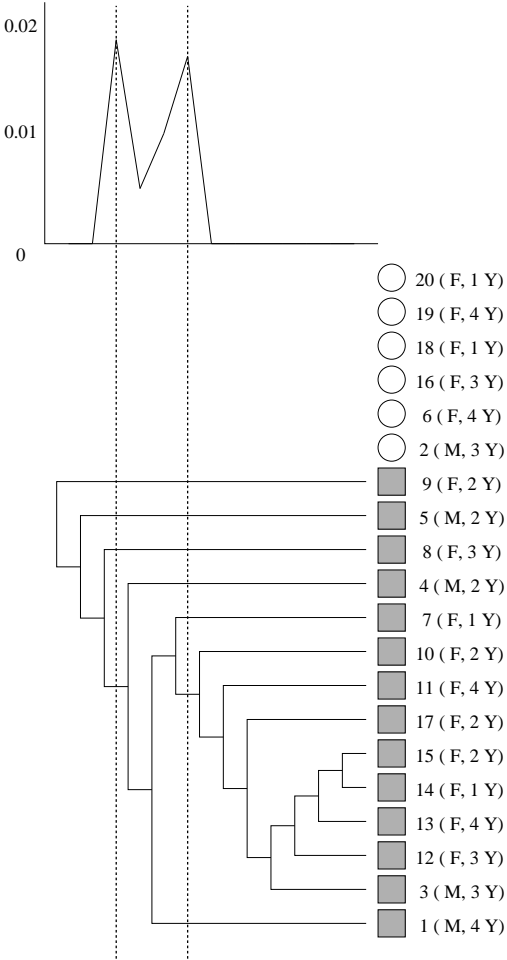,angle=90,width=8cm}
    \caption{\label{monkeys_eff}{Dendrogram of the primate network.
The circles represent the asocial monkeys, the
squares the social monkeys (see text).
There is no separation in classes; our procedure leads to 
a progressive isolation of the nodes. The modularity $Q$ is very low, the higher
peak is relative to a partition in a large group and the isolated nodes
$5$, $8$ and $9$ (besides the asocial primates).}}
  \end{center}
\end{figure}

In this section we consider a data set collected by Linda 
Wolfe \citep{lmcentrality,groups}, recording 3 months of interactions 
amongst a group of 20 monkeys, where interactions were defined as 
the joint presence at the river. 
The dataset also contains information on the sex and the age 
of each animal. Monkeys 1-5 are males, monkeys 6-20 are females. 
In increasing order of age: monkeys 7, 14, 18, 20 belong to the first 
age group (the youngest), monkeys 4, 5, 9, 10, 15, 17 to the second, 
monkeys 2, 3, 8, 12, 16 to the third and monkeys 1, 6, 11, 13, 19 to the fourth 
and oldest group. A detailed analysis of the individual and group 
centrality of this network can be found in Refs. \cite{lmcentrality,groups}.  
The total number of links
is 31, i.e. of the order of magnitude of the nodes.
Indeed, six out of twenty monkeys did not actively participate
in the social life of the group; 
the resulting non-directed non-valued 
graph thus consists of 6 isolated points (labeled by the numbers
2, 6, 16, 18, 19, 20) and a connected component of 14 points. 
The results of our analysis are illustrated in Fig. \ref{monkeys_eff},
where we reported as well for each primate both sex (M=male, F=female) and age
(in years).
The modularity of the subsequent subdivisions of the network in
components is very low, which shows that there is no appreciable
community structure; 
nevertheless, two peaks are clearly visible, 
the higher of which is obtained when the nodes
5, 8 and 9 separate one after the other from the network. One gets then a 
major community of eleven elements and nine isolated monkeys.
We do not find any sensible relationships between our partition and
the division of the primates in age groups. 
We analyzed the network as well with the method of Girvan and Newman
and the results are essentially the same: one gets again two peaks for the
modularity (whose values remain low) 
and the best partition of the network corresponds to a separation
in the same large community we found before without node 11, which is 
now isolated, plus isolated sites except the pair 5-8.

\section{Conclusions}
\label{final}

We have presented a new algorithm to identify the 
subdivisions of complex networks in cohesive groups of vertices, or
communities. 
The algorithm is based on a recently introduced centrality measure,
the so-called information centrality, and consists in 
classifying all edges according to the value of this measure, so to 
determine which edge is most central: the latter edge is then removed
from the network. 
One then recalculates the information centrality of the remaining
edges and again removes the most central edge; the procedure 
is repeated until all
edges are removed.
The hope is that this sequential removal of edges looses the bonds between 
tightly connected groups of vertices,
so that, at some stage, they eventually separate from each other.

For the quantitative evaluation of the goodness of the successive splits, 
which is necessary in order to identity the best subdivision of the network,
we adopted the modularity $Q$ introduced in \cite{newgirv2}. 
Our algorithm runs to completion in time $O(K^3N)$ 
($K$ and $N$ are the number of edges and vertices of
the graph, respectively)
and therefore is
not so fast as other methods; because of that, networks with 
thousands of vertices are unreachable. 
The aim of the paper, however, was to 
check whether the information centrality is relevant in the search 
of the communities. 

The results of the application of our method both to 
computer generated networks and to real networks clearly show that the 
algorithm is indeed able to detect the real communities in most cases.
This implies the existence of a correlation between the
information centrality ${C^I}_k$ of an edge $k$ and the fact that
the edge joins two different communities; the higher ${C^I}_k$, the more likely
$k$ is a tie between groups. 
This is confirmed by the correlation we observed between the peaks of 
$C^I$ and the jumps in the modularity (see Figs. \ref{randeff} and \ref{karateeff}).
We stressed the importance of the recalculation of the information centrality
step by step; without it the algorithm is not able to distinguish the communities. 
Our method was especially devised for sparse graphs (i.e. when $K{\sim}N$), and it 
is probably doomed to fail for dense graphs ($K{\sim}N^2$).

The examples we have taken allowed us as well to see how efficient our algorithm is 
compared with others. In particular we made extensive comparisons with the
algorithm of Girvan and Newman \cite{newgirv1,newgirv2}, which also uses 
a centrality measure, the edge betweenness. 
It turns out that our algorithm is generally as good as the one of Girvan and
Newman.
It seems to perform slightly 
better when there is a high degree of mixture between the classes;
on the other hand, it sometimes has troubles with nodes which are too loosely
bound to the rest of the network (like nodes with a single edge), which may
separate too early and be misclassified, although they often happen to be
truly independent communities.

{\bf Acknowledgements } 
We thank B. A. Huberman for providing us with the network 
of the college football teams and G. Caldarelli for
useful information on the food webs. S. F. acknowledges the support of 
the DFG Forschergruppe under grant FOR 339/2-1.


\begin{thebibliography}{99}
\bibitem{note1}
Newman and Girvan in Ref.\cite{newgirv2} 
have also introduced a third measure of edge betweenness, 
the current-flow betweenness, defined by considering 
an electric circuit associated to the graph 
and by using Kirchoff's laws. Despite the different 
definition they have proven that 
the current-flow betweenness coincides with the  
random-walk betweenness.
\bibitem{note2} 
Newman and Girvan consider all edges in the original network, 
even after edges have been removed by the community structure 
algorithm the modularity is
calculated by using the full network.
\end{thebibliography}
\end{document}